\documentclass[12pt]{article} 

\usepackage{amsmath}
\usepackage{graphicx}
\usepackage{amsfonts}
\usepackage{biblatex}
\usepackage{hyperref} 
\newcommand{\myemail}[1]{\href{mailto:#1}{#1}}

\title{A mathematical model for pricing perishable goods for quick-commerce applications \\
  \small Concept Note
} %

\author{Milon Bhattacharya \\ 
    \emph{Indian Institute of Management Visakhapatnam} \\
    \myemail{milon.bhattacharya25-08@iimv.ac.in} 
}

\date{\today} 

\begin{document} 

\maketitle 

\section*{Introduction} 
Quick commerce (q-commerce) is one of the fastest growing sectors in India. It provides informal employment to approximately 4,50,000 workers, and it is estimated to become a USD 200 Billion industry \cite{ranjekar2023rise} by 2026. A significant portion of this industry deals with perishable goods. (e.g. milk, dosa batter etc.) These are food items which are consumed relatively fresh by the consumers and therefore their order volume is high and repetitive even when the average basket size is relatively small \cite{jirapatsil2022market}. The fundamental challenge for the retailer is that, increasing selling price would hamper sales and would lead to unsold inventory. On the other hand setting a price less, would lead to forgoing of potential revenue. This paper attempts to propose a mathematical model which formalizes this dilemma. The problem statement is not only important for improving the unit economics of the perennially loss making quick commerce firms, but also would lead to a trickle-down effect in improving the conditions of the gig workers as observed in \cite{kalbalia2025regulating}. The sections below describe the mathematical formulation. The results from the simulation would be published in a follow-up study.

\section*{Problem Statement}

The problem of dynamic pricing has been studied well in the last few decades. We borrow copiously from the pioneering works of \cite{minga2003dynamic} , \cite{guo2025dynamic} and \cite{chornous2020modeling} which attempt to tackle the problems using either using rigorous formulation, simulation or simply using a supervised learning based model.

\subsection*{Mathematical Notation}
We begin the description of the problem by describing the mathematical notation used in the following sections. 

For simplicity’s sake, let's consider a set of $N$ perishable SKUs sold over a discrete time horizon $t = 1, \dots, K$ on a quick commerce platform. Each SKU $i \in \{1, \dots, N\}$ has inventory $I_{i,t}$, selling price $p_{i,t}$, unit cost $c_i$, and platform commission $\eta$. Unsold units incur a terminal salvage penalty $s_i$.

The formulation assumes that the simulation starts at $t = 0 $. It will go upto a maximum of $t = T$. The time is not discrete but a ratio variable. 
\begin{equation}
     t (time) \in [0, T]
\end{equation}

The price function (which has to be estimated) output an optimal value of product price  which is always greater than zero: 
\begin{equation}
  \mathbf{p}(t) \ge 0
\end{equation}
instantaneous demand rate (units per unit time) when price is $p$ at time $t$ is given by:
\begin{equation}
D(t, p)
\end{equation}

The inventory held for the SKU \emph{s} at time $t$. The model assumes replenishment at $t=0$

\begin{equation}
  I(t) ; I(0) = I_0
\end{equation}

$h$ is the per-unit terminal cost of unsold item at $T$ (disposal or salvage cost; can be negative if salvage yields revenue).

Further, $c$: is the marginal cost of the item; if present profit uses $p(t)-c$ rather than $p(t)$. This will be non-zero for items which require costs in addition to delivery costs.

$DI(t)$ is the function which captures the baseline time-varying demand intensity (freshness decay, time-of-day). It is a function of product features, marketing spend etc.

\subsection*{Assumptions}
\begin{itemize}
  \item The customers orders are received regularly which can be estimated using a Poisson distribution. The probability of receiving an order is given by $P(X)$
  
  \begin{equation}
    P(X = k) = \frac{\lambda^k e^{-\lambda}}{k!}
  \end{equation} 

Where:
\begin{itemize}
    \item $P(X = k)$: Probability of the platform receiving $k$  orders at a given point in time.
    \item $\lambda$: The rate parameter (average number of order received per unit of time). $\lambda > 0$.
    \item $e$: Euler's number ($e \approx 2.71828$).
    \item $k$: The number of occurrences ($k \in \{0, 1, 2, \dots\}$).
    \item $k!$: The factorial of $k$.
\end{itemize}

In subsequent studies, our attempt would be to guess the distribution  $P(X)$. For, completeness sake Figure \ref{fig:orderday} and Figure \ref{fig:orderweek} gives a rough distribution of order received across different times in a day and week respectively. As it can be observed from the illustrations, the demand per unit of time (let's say minutes) is stable in the short term. This allows us to estimate $\lambda$ for a given time of the day. Also, unlike previous attempts of modeling this parameter as a distribution, this approach is much more data-driven. 

\begin{figure}[h!] 
    \centering 
    \includegraphics[width=0.8\textwidth]{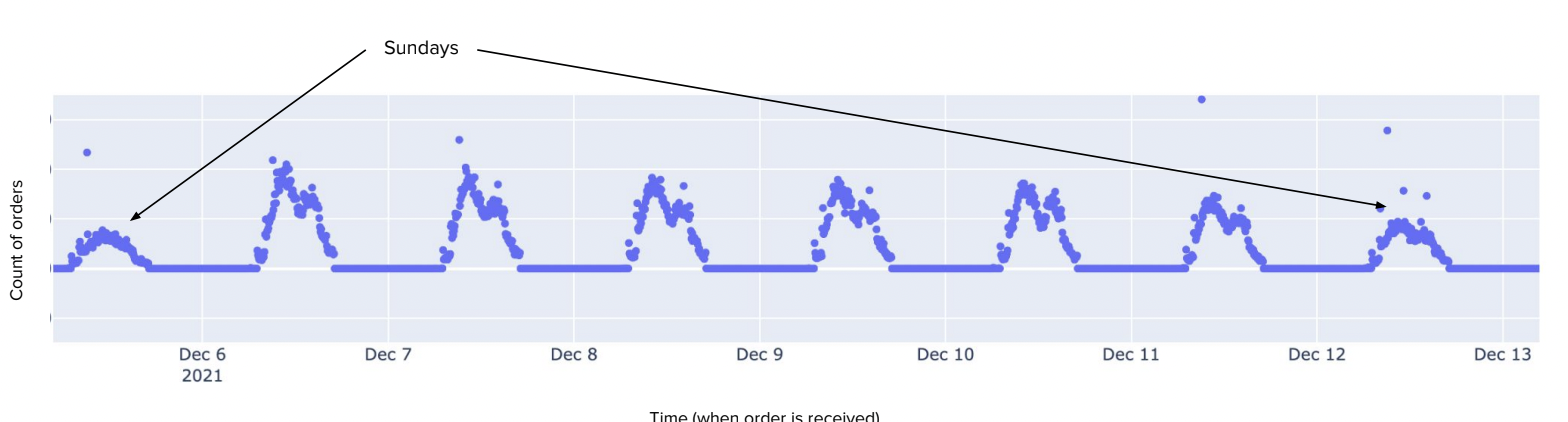} 
    \caption{Aggregated orders received across a week (window of 60 minutes)} 
    \label{fig:orderweek} 
\end{figure}

\begin{figure}[h!] 
    \centering 
    \includegraphics[width=0.8\textwidth]{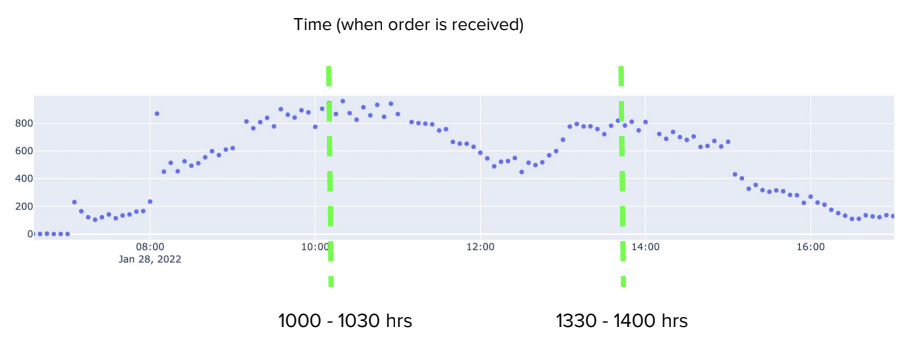} 
    \caption{Aggregated orders received across a day (window of 5 minutes)} 
    \label{fig:orderday} 
\end{figure}

  \item Each order the customer has an option of ordering any combination of the available SKUs depending on the attributes of the SKU/item (which \emph{excludes} the price for our current formulation); that choice can be a single SKU, multiple SKUs (bundles / complements), or none. 
  
  \item As mentioned before customer orders are stochastic counting processes constrained by inventory. Platform posts a time-varying price vector $\mathbf{p}(t)$; platform takes commission fraction $\phi$ (which can constant or SKU-specific).
  \item Inventory of each SKU perishes at horizon $T$ (or earlier)---unsold units incur salvage/disposal cost.
  \item The aim of the analysis is to choose price path $\mathbf{p}(t)$ to maximize expected profit as order are received / fulfilled and inventory decreases.
\end{itemize}

\section*{Customer Orders}

As was discussed in previous sections, customers arrive according to a Poisson process with rate $\lambda$. Each arrival chooses a basket $b \subseteq A_t$, where $A_t = \{ i \mid I_{i,t} > 0 \}$ denotes available SKUs. The utility of basket 
$b$ is given by:
\begin{equation}
    U_b(p_t) = \sum_{i \in b} (\alpha_i - \beta_i p_{i,t}) + \gamma \max(0, |b| - 1),
\end{equation}
where $\gamma$ captures complementarity ($\gamma > 0$) or substitutability ($\gamma < 0$) among items.
The probability of a customer choosing basket $b$ follows a multinomial logit structure:
\begin{equation}
    P(b \mid p_t, A_t) = 
    \frac{\exp(U_b(p_t)) \cdot 1_{\{b \subseteq A_t\}}}
    {\sum_{b' \subseteq A_t} \exp(U_{b'}(p_t))}.
\end{equation}
\subsection*{Demand and Inventory Dynamics}

The expected demand for SKU $i$ in epoch $t$ is:
\begin{equation}
    D_i(p_t) = \lambda \, \Delta t \sum_{b \ni i} P(b \mid p_t, A_t).
\end{equation}
Inventory evolves as:
\begin{equation}
    I_{i,t+1} = \max(0, I_{i,t} - D_i(p_t) + \xi_{i,t}),
\end{equation}
where $\xi_{i,t}$ captures stochastic variation in realized demand.

\subsection*{Expected Profit}
Expected profit for epoch $t$ is:
\begin{equation}
    \Pi_t(p_t, I_t) = 
    \sum_{i=1}^N D_i(p_t) \big[ (1 - \eta)p_{i,t} - c_i \big],
\end{equation}
with terminal salvage loss:
\begin{equation}
    \Pi_T(I_T) = - \sum_{i=1}^N s_i I_{i,T}.
\end{equation}
The optimization objective is to maximize the expected total profit:
\begin{equation}
    \max_{p_t,\, t=1..K} \; 
    \mathbb{E}\Bigg[\sum_{t=1}^{K} \Pi_t(p_t, I_t) + \Pi_T(I_T)\Bigg].
\end{equation}

\section*{Guardrails on inventory}
To reduce the risk of stockouts, we define a safety-adjusted target demand:
\begin{equation}
    \tilde{D}_{i,t} = \rho \, \frac{I_{i,t}}{K - t + 1}, 
    \quad \emph{where } 0 < \rho < 1.
\end{equation}

Prices are chosen such that the expected demand approximates the target:
\begin{equation}
    D_i(p_t) \approx \tilde{D}_{i,t},
\end{equation}
using stochastic approximation or root-finding.
To avoid excessive volatility, prices satisfy an inertia constraint:
\begin{equation}
    |p_{i,t} - p_{i,t-1}| \leq \Delta_{\max}.
\end{equation}

\subsection*{Inventory pricing}
We approximate the value of remaining inventory using a linear value function:
\begin{equation}
    V_t(I_t) \approx w_t^\top I_t,
\end{equation}
where $w_t$ is updated by fitted value iteration:
\begin{equation}
    w_t = \arg\min_w 
    \mathbb{E}\big[(r_t + V_{t+1}(I_{t+1}) - w^\top I_t)^2\big].
\end{equation}
The resulting policy chooses prices to maximize current and expected future value:
\begin{equation}
    p_t^* = \arg\max_{p_t} 
    \Big[
        \Pi_t(p_t, I_t) + 
        \mathbb{E}\big[V_{t+1}(I_{t+1}) \mid p_t, I_t\big]
    \Big].
\end{equation}

\section*{Computation of demand}
\subsection*{Derivation of Expected Demand}

Let $\mathcal{B}$ denote the power set of available SKUs.  
For each basket $b \in \mathcal{B}$, define the choice probability:

\begin{equation}
  P(b) = \frac{\exp(U_b(p_t))}{\sum_{b' \in \mathcal{B}} \exp(U_{b'}(p_t))}
\end{equation}

Then the expected demand for item $i$ is:
\begin{equation}
  d_i(p_t) = \sum_{b \ni i} P(b)
\end{equation}

For Poisson arrivals with rate $\lambda$, the expected number of units sold in period $\Delta t$ is:
\begin{equation}
  D_i(p_t) = \lambda \, \Delta t \, d_i(p_t)
\end{equation}

\subsection*{Modeling of purchases}
Realized sales $\tilde{D}_{i,t}$ are modeled as random variables:

\begin{equation}
  \tilde{D}_{i,t} = D_i(p_t) + \epsilon_{i,t},
\end{equation}

where $\epsilon_{i,t}$ is mean-zero noise.

Inventory updates recursively as:
\begin{equation}
  I_{i,t+1} = \max(0, I_{i,t} - \tilde{D}_{i,t})
\end{equation}

\subsection*{Estimating }
To hedge against uncertainty, we define a sales target proportional to remaining stock:
\begin{equation}
  \tilde{D}_{i,t} = \rho \frac{I_{i,t}}{K - t + 1}, \quad \rho < 1.
\end{equation}

Prices are determined numerically by solving:
\begin{equation}
  D_i(p_t) - \tilde{D}_{i,t} = 0
\end{equation}

This can be solved either using approximations or gradient-based approaches.

\subsection*{Preventing sudden changes to price}
To ensure temporal consistency in prices, we enforce:
\begin{equation}
  |p_{i,t} - p_{i,t-1}| \leq \Delta_{\max}
\end{equation}

This prevents abrupt price shifts that can confuse customers or violate pricing policies.

\subsection*{Value Function Approximation and ADP}
We approximate the continuation value $V_t(I_t)$ by a linear function in the inventory:

\begin{equation}
  V_t(I_t) \approx w_t^\top I_t
\end{equation}
Given sampled trajectories $(I_t, p_t, r_t, I_{t+1})$, we estimate $w_t$ by least-squares regression:
\begin{equation}
  w_t = \arg\min_w \mathbb{E}\left[ \big( r_t + V_{t+1}(I_{t+1}) - w^\top I_t \big)^2 \right]
\end{equation}

The corresponding greedy policy:
\begin{equation}
p_t^* = \arg\max_{p_t}
\left[
    \Pi_t(p_t, I_t) +
    \mathbb{E}\big[V_{t+1}(I_{t+1}) \mid p_t, I_t\big]
\right].
\end{equation}
This policy effectively balances immediate expected profit with the future shadow value of remaining inventory.
\subsection*{Final Algorithm}

\begin{enumerate}
    \item Initialize $I_0$, $p_0$, and $V_T(I_T) = -\sum_i s_i I_{i,T}$.
    \item For each epoch $t = 1,\dots,K$:
    \begin{enumerate}
        \item Compute safe target demand $\tilde{D}_{i,t}$.
        \item Solve $D_i(p_t) \approx \tilde{D}_{i,t}$ using root-finding.
        \item Enforce price inertia constraint.
        \item Update inventory $I_{i,t+1}$ via stochastic dynamics.
    \end{enumerate}
    \item Update $w_t$ using fitted value iteration.
\end{enumerate}

\printbibliography
\end{document}